\documentclass[10pt, conference,a4paper]{IEEEtran}
\usepackage[left=1.5cm,right=1.5cm,top=1.9cm, bottom=1.9cm]{geometry}
\IEEEoverridecommandlockouts
\usepackage{cite}
\usepackage{amsmath,amssymb,amsfonts}
\usepackage{algorithmic}
\usepackage{graphicx}
\usepackage{textcomp}
\usepackage{xcolor}
\usepackage{listings}
\usepackage{pdfpages}
\usepackage{bm}
\usepackage{textcomp}
\usepackage{algorithm2e}
\usepackage{balance}
\usepackage{flushend}

\SetAlFnt{\small}
\SetAlCapFnt{\small}
\SetAlgoVlined
\RestyleAlgo{ruled}
\SetKwComment{Comment}{\# }{}
\SetKw{init}{\textbf{Initialization :}}

\def\BibTeX{{\rm B\kern-.05em{\sc i\kern-.025em b}\kern-.08em
    T\kern-.1667em\lower.7ex\hbox{E}\kern-.125emX}}

\usepackage{tikz}
\usepackage{circuitikz}
\usetikzlibrary{positioning}
\usepackage{pgfkeys}

\usetikzlibrary{patterns}

\usepackage{pgfplots}
\pgfplotsset{compat=newest}
\pgfplotsset{plot coordinates/math parser=false}
\newlength\figureheight
\newlength\figurewidth

\usetikzlibrary{positioning}

\usetikzlibrary{arrows.meta,positioning}
\tikzset{block/.style={draw, rectangle, fill=cyan!90,
        minimum height=2em, minimum width=3em},
    sum/.style={draw, circle, node distance=1cm},
    input/.style={coordinate},
    output/.style={coordinate},
    pinstyle/.style={pin edge={to-,thin,black}},
        saturation block/.style={%
            draw,
            path picture={
                \pgfpointdiff{\pgfpointanchor{path picture bounding box}{south west}}%
                {\pgfpointanchor{path picture bounding box}{north east}}
                \pgfgetlastxy\x\y
                \tikzset{x=\x*.4, y=\y*.4}
                \draw [very thin] (-1,0) -- (1,0) (0,-1) -- (0,1);
                \draw [very thick] (-1,-.7) -- (-.7,-.7) -- (.7,.7) -- (1,.7);
            },
        }
    }

\tikzset{%
        rateLimit block/.style={%
            draw,
            path picture={
                \pgfpointdiff{\pgfpointanchor{path picture bounding box}{south west}}%
                {\pgfpointanchor{path picture bounding box}{north east}}
                \pgfgetlastxy\x\y
                \tikzset{x=\x*.4, y=\y*.4}
                \draw [very thin] (-1,0) -- (1,0) (0,-1) -- (0,1);
                \draw [very thick] (-1,-1) -- (1, 1);
            },
        }
    }
    \usetikzlibrary{svg.path}
    \usepackage{scalerel}

    \definecolor{orcidlogocol}{HTML}{A6CE39}
    \tikzset{
      orcidlogo/.pic={
        \fill[orcidlogocol] svg{M256,128c0,70.7-57.3,128-128,128C57.3,256,0,198.7,0,128C0,57.3,57.3,0,128,0C198.7,0,256,57.3,256,128z};
        \fill[white] svg{M86.3,186.2H70.9V79.1h15.4v48.4V186.2z}
                     svg{M108.9,79.1h41.6c39.6,0,57,28.3,57,53.6c0,27.5-21.5,53.6-56.8,53.6h-41.8V79.1z M124.3,172.4h24.5c34.9,0,42.9-26.5,42.9-39.7c0-21.5-13.7-39.7-43.7-39.7h-23.7V172.4z}
                     svg{M88.7,56.8c0,5.5-4.5,10.1-10.1,10.1c-5.6,0-10.1-4.6-10.1-10.1c0-5.6,4.5-10.1,10.1-10.1C84.2,46.7,88.7,51.3,88.7,56.8z};
      }
    }

    \newcommand\orcidicon[1]{\href{https://orcid.org/#1}{\mbox{\scalerel*{
    \begin{tikzpicture}[yscale=-1,transform shape]
    \pic{orcidlogo};
    \end{tikzpicture}
    }{|}}}}
\usepackage[hidelinks]{hyperref}

\makeatother

\usepackage{eso-pic}

\makeatletter

\let\old@ps@IEEEtitlepagestyle\ps@IEEEtitlepagestyle
\def\confheader#1{%
    \def\ps@IEEEtitlepagestyle{%
        \old@ps@IEEEtitlepagestyle%
        \def\@oddhead{\strut\hfill#1\hfill\strut}%
        \def\@evenhead{\strut\hfill#1\hfill\strut}%
    }%
    \ps@headings%
}

\makeatother

\begin{document}

% \title{ Codes Design and  Performance Enhancement\\ of Short Block-Length Uplink Channels}
%\title{ Codes Design and Performance Enhancement\\ for Short Block-Length Uplink Channels}

\title{ Low-complexity Block-Based Decoding \\Algorithms for Short Block Channels}
% \author{\IEEEauthorblockN{ Mody~Sy~\IEEEmembership{}\orcidicon{0000-0003-2841-2181} ~%Student Member,~IEEE,
%         and~ Raymond~Knopp~\IEEEmembership{}\orcidicon{0000-0002-6133-5651}}
%
%         \IEEEauthorblockA{\textit{Communication Systems Dept.}\\ \textit{EURECOM}, 06410 BIOT, France \\
%         Email: mody.sy@eurecom.fr,~raymond.knopp@eurecom.fr}
% }
%\conf{ 2023 IEEE AFRICON}
% ===== Conference Name =====
  \markboth{\textbf{2023 IEEE AFRICON}}{}
% \author{\IEEEauthorblockN{ Mody~Sy~\IEEEmembership{} \orcidicon{0000-0003-2841-2181} and~ Raymond~Knopp~\IEEEmembership{}\orcidicon{0000-0002-6133-5651} }%Student Member,~IEEE,
%         \IEEEauthorblockA{
%         %\textit{Dept. of Communication Systems, EURECOM }\\
%         \textit{EURECOM }, 06410 BIOT, France \\
%         Email: mody.sy@eurecom.fr, ~raymond.knopp@eurecom.fr}
% }

\author{\IEEEauthorblockN{ Mody~Sy$\textsuperscript{\orcidicon{0000-0003-2841-2181}}$\IEEEmembership{}~and~ Raymond~Knopp$\textsuperscript{\orcidicon{0000-0002-6133-5651}}$}

       \IEEEauthorblockA{
      %\textit{Communication Systems Dept.}\\
        \text{EURECOM}, 06410 BIOT, France \\
        \textcolor{darkgray}{mody.sy@eurecom.fr,~raymond.knopp@eurecom.fr}}
}
\maketitle
\begin{abstract}
This paper presents low-complexity block-based encoding and decoding algorithms for short block length channels. In terms of the precise use-case, we are primarily concerned with the baseline 3GPP Short block transmissions in which payloads are encoded by Reed-Muller codes and  paired with orthogonal DMRS. In contemporary communication systems, the short block decoding often employs the utilization of DMRS-based least squares channel estimation, followed by maximum likelihood decoding. However, this methodology can incur substantial computational complexity when processing long bit length codes. We propose an innovative approach to tackle this challenge by introducing the principle of block/segment encoding using First-Order RM Codes which  is amenable to low-cost decoding through block-based fast Hadamard transforms. The Block-based FHT has demonstrated to be cost-efficient with regards to decoding time, as it evolves from quadric to quasi-linear complexity with a manageable decline in performance. Additionally, by incorporating an adaptive DMRS/data power adjustment technique, we can bridge/reduce the performance gap and attain high sensitivity, leading to a good trade-off between performance and complexity to efficiently handle small payloads.
\end{abstract}

\section{Introduction}
5G NR aims to effectively transmit small payloads, typically consisting of tens of bits, with low error rates even in low signal-to-noise environments. This requires the design of strong structured coding strategies and low-complexity decoding algorithms to meet the demanding requirements of URLLC use cases.
In this paper, we explore potential methods for improving the detection and decoding performance with low computational complexity in 5G/6G systems. Our focus is on the use-case of 3GPP short block transmissions paired with orthogonal {\em Demodulation Reference Signals} or DMRS, which employ Reed-Muller coding schemes. Thus, the DMRS symbols, which are message-independent, are utilized by the gNodeB receiver for resolving channel uncertainty through explicit channel estimation or joint estimation and detection. In 3GPP systems, short blocks, ranging from 3 to 11 bits, are encoded using a family of Reed-Muller codes of dimension ${\mathcal C}(32, K)$ prior to being transmitted on the uplink control channels \cite{3GPP38212}. On reception, the messages are recovered using maximum likelihood decoding, which can be computationally demanding.\\
Reed-Muller codes (RM codes) are commonly known to be decodable using {\em Hadamard Transform} (HT) or {\em Fast Hadamard Transform} (FHT). However, it is well-established that decoding the first-order RM code ($RM(r=1, m)$) using FHT is easier compared to higher-order RM codes ($r\geq 2$). In recent literature, several innovative algorithms have been proposed for decoding RM codes of any order \cite{Dumer2006, Ye2019, Li2021, Lian2020, Doan2022}.\\
Although maximum likelihood decoding algorithms have been extensively investigated in traditional literature for decoding data packets encoded with first-order Reed-Muller codes \cite{Ashikhmin}, it can become computationally expensive when the message length exceeds 6 bits. This is because the resulting codewords tend to be excessively long, leading to complex decoding processes that involve high-dimensional Hadamard transforms when using a FHT-based decoder.
This presents a significant challenge for transmitting short packets as the cost can be substantial.
As an illustration, for a message of $K=11$ bits, the length of the code words would be $N^\prime=2^{10}$ bits using a first order $RM(1, m=10)$. Hence, to address the constraint of having a message length of $K\geq 6$ bits, we can utilize the principle of encoding and decoding by blocks. This method takes advantage of the low complexity decoding offered by FHT-based decoders. \\ The objective is to segment the message into smaller, more manageable segments of bits, which can then be fed into $RM(1,m)$ encoders and concatenated. Upon reception, the received code is deconcatenated and decoded through the appropriate dimension of the Hadamard Transform which is amenable to a low complexity receiver.\\
The manuscript is structured as follows. Section II lays out the system model foundation, Section III highlights the proposed encoding/decoding methods, Section IV presents the results and performance analysis, and finally Section V concludes the paper.
\section{System Model }
Consider a discrete-time model in which the transmitted and received symbols are $N$-dimensional column vectors, and thus a system is designed in such a way that the relationship between the transmitted and received signals is as follows:
\begin{equation}
\centering
    %\mathbf {y} = \diag{\mathbf {h}}\mathbf {x}+ \mathbf {z}
    \mathbf {y} = \mathbf {h}\mathbf {x}+ \mathbf {z},
\end{equation}
where $\mathbf {y}$  represents an observed vector in $N$ complex dimensions, $\mathbf {x}$  is a $N$-dimensional modulated vector,  $z$ is additive white Gaussian noise whose the real and imaginary components are independent and have variance $\sigma^2$.
%The performance of the system is predicted using 3GPP channel models.
The transmitted vector is composed of data-independent components, known as pilot or reference signals, that help resolve channel ambiguity in time, frequency, and space. These reference signals are used to estimate the vector channels $\mathbf {h}$ and are {\em interleaved} with data-dependent components  in accordance with the attributes of the propagation channel.
\begin{equation}
\centering
    \mathbf {x} =\mathbf {x}^{(d)}+ \mathbf {x}^{(p)}
\end{equation}
where subscripts $(d)$ and $(p)$ serve to denote the data components  and reference signals respectively.
The number of data dimensions is denoted by $N_d$, and the number of reference signal dimensions is denoted by $N_p$, where $N_d+N_p=N$.
In 3GPP, the standard notation for $N$ is $12\mathcal P \mathcal L$, where $\mathcal P$ refers to the number of physical resource blocks, each consisting of $12$ complex dimensions or resource elements. The value of $\mathcal P$ usually falls within a range of $1$ to $16$. $\mathcal L$ represents the number of symbols, typically ranging from $1$ to $14$, but it can be increased if multiple slots are utilized to signal the channel bits.
\section{Block-based Channel Coding For Short Data}
\subsection{Block-based Encoding Principle}
% The Reed-Muller codes were pioneers among codes utilized in space communication % % % systems. They were one of the earliest codes to offer a means of achieving a % % % specified minimum distance, making them highly attractive to engineers and scientists alike. This, along with the presence of a fast maximum likelihood decoding algorithm, has contributed to their continued popularity and usage.
% These binary codes are denoted as $RM(r,m)$ where $r$ and $m$ represent the order and variables.
%\subsubsection{ Encoding and Decoding Algorithms for First-Order RM Codes}

% We describe algorithms for encoding  $RM(1,m)$ codes, which are $(2^m , m + 1, 2^{m-1})$ codes.
In instances where the payload exceeds $6$ bits, such as in the case of $K = 11$ bits, a combination of two first-order Reed-Muller codes, $RM(1, m=4)$ and $RM(1, m=5)$, can be employed to encode the respective sub-blocks of 5 bits and 6 bits.\\
In regards to the Reed-Muller code $RM(1, 4)$, the codewords are generated using (\ref{eqn:enc_sblk1}).
\begin{equation}\label{eqn:enc_sblk1}
\mathbf {c}^{(1)}=\mathbf {m}^{(1)}{\mathbf  G}^{(1)}= \mathbf  m^{(1)} \left[\begin{array}{c}
% \mathbf {1} \\
% \mathbf {v}_4 \\
% \mathbf {v}_3 \\
% \mathbf {v}_2 \\
% \mathbf {v}_1
\mathbf {1}\quad\mathbf {v}_4\quad\mathbf {v}_3\quad\mathbf {v}_2\quad\mathbf {v}_1
\end{array}\right]^\mathrm T,
\end{equation}
% where
% \begin{equation}
%   \begin{aligned}
%     \mathbf  m^{(1)}&= (m^{(1)}_0, m^{(1)}_4, m^{(1)}_3, m^{(1)}_2, m^{(1)}_1),\\
%     \mathbf  c^{(1)} &= (c^{(1)}_0, c^{(1)}_1, \ldots,  c^{(1)}_{15}).
%   \end{aligned}\nonumber
% \end{equation}
This code is characterized as an $(N^\prime=16, K=5, d_{min}=8)$ code, where the minimum distance of the $RM(r,m)$ is defined as $2^{m-r}$. The monomials of degree less than or equal to $r$ are represented by ${\mathbf  1, \mathbf   v_1, \mathbf  v_2, \mathbf  v_3, \mathbf  v_4}$, with associated vectors.
\begin{table}[htbp]
  \centering
  %\caption{Add caption}
    \begin{tabular}{|r|r|r|r|r|}
    \hline
    {$\mathbf  1$} & \multicolumn{1}{l|}{{$\mathbf  v_4$}} & \multicolumn{1}{l|}{{$\mathbf  v_3$}} & \multicolumn{1}{l|}{{$\mathbf  v_2$}} & \multicolumn{1}{l|}{{$\mathbf  v_1$}} \\
    \hline
    \hline
    1     & 0     & 0     & 0     & 0 \\
    \hline
    1     & 0     & 0     & 0     & 1 \\
    \hline
    1     & 0     & 0     & 1     & 0 \\
    \hline
    1     & 0     & 0     & 1     & 1 \\
    \hline
    1     & 0     & 1     & 0     & 0 \\
    \hline
    1     & 0     & 1     & 0     & 1 \\
    \hline
    1     & 0     & 1     & 1     & 0 \\
    \hline
    1     & 0     & 1     & 1     & 1 \\
    \hline
    1     & 1     & 0     & 0     & 0 \\
    \hline
    1     & 1     & 0     & 0     & 1 \\
    \hline
    1     & 1     & 0     & 1     & 0 \\
    \hline
    1     & 1     & 0     & 1     & 1 \\
    \hline
    1     & 1     & 1     & 0     & 0 \\
    \hline
    1     & 1     & 1     & 0     & 1 \\
    \hline
    1     & 1     & 1     & 1     & 0 \\
    \hline
    1     & 1     & 1     & 1     & 1 \\
    \hline
    \end{tabular}%
  \label{tab:addlabel}%
\end{table}%

If we consider $RM(1, 5)$,  code words are generated by
\begin{equation}
\mathbf {c}^{(2)}=\mathbf {m}^{(2)}{\mathbf  G}^{(2)}= \mathbf  m^{(2)} \left[\begin{array}{c}
% \mathbf {1} \\
% \mathbf {v}_5 \\
% \mathbf {v}_4 \\
% \mathbf {v}_3 \\
% \mathbf {v}_2 \\
% \mathbf {v}_1
\mathbf {1}\quad\mathbf {v}_5\quad\mathbf {v}_4\quad\mathbf {v}_3\quad\mathbf {v}_2\quad\mathbf {v}_1
\end{array}\right]^\mathrm T .\nonumber
\end{equation}
% where
% \begin{equation}
%   \begin{aligned}
%     \mathbf  m^{(2)} &= (m^{(2)}_0, m^{(2)}_5, m^{(2)}_4, m^{(2)}_3, m^{(2)}_2,m^{(2)}_1),\\
%     \mathbf  c^{(2)} &= (c^{(2)}_0, c^{(2)}_1, \ldots,  c^{(2)}_{31})
%   \end{aligned}\nonumber
% \end{equation}
This is a $(N^\prime=32, K=6, d_{min}=16)$ code.

Therefore, the process of concatenation entails the merging of the two sub-codes

\begin{equation}
\mathbf  c=[\mathbf  c^{(1)} \ \mathbf  c^{(2)}] = [\mathbf  m^{(1)}\mathbf  G^{(1)} \ \mathbf  m^{(2)}\mathbf  G^{(2)} ].
\end{equation}
The resultant bit sequence, prior to the concatenation of code blocks, is represented as $c(0), \ c(1), \ c(2), \ \ldots, \ c(N^\prime-1)$. The output bit sequence, following rate matching, is denoted as $e(0), \ e(1), \ e(2), \ \ldots, \ e(E-1)$, where the length of the rate matching output sequence, $E$, is dependent on the number of {\em physical resource blocks} or (PRBs). This information block, represented as $\mathbf  e$, is then subjected to  {\em quadrature phase shift keying} (QPSK) modulation, resulting in a block of complex-valued modulation symbols $x(0), \ x(1), \ \ldots, \ x\left(E / 2-1\right)$. The resource mapping process follows, whose aim is to allocate the modulated symbols onto resource occasions.
%as depicted in Figure~\ref{fig:re_mapp_pucch2_1symb1}.

\begin{figure*}[!ht]
        \centering
        \includegraphics[width=0.9\linewidth]{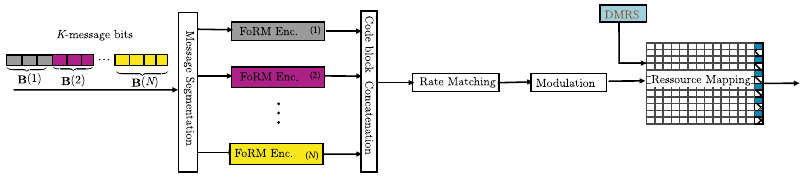}
        \caption{ Block-based RM(1, M) encoding of Short block -length :Transmitter end.}
       \label{fig:rmfirt_order_tx}
\end{figure*}

% \begin{figure}[ht]
%   \centering
%   \includegraphics[width=0.55\linewidth]{figures/tikz/ray/4dmrs.pdf}
%   \caption{{General resource mapping}}
%   \label{fig:re_mapp_pucch2_1symb1}
% \end{figure}
The procedure of block-based RM encoding is depicted in Figure~\ref{fig:rmfirt_order_tx}.
\subsection{Block-based Decoding via FHT}%\subsubsection*{Hadamard Matrices}~\\
A Hadamard matrix, represented as $\mathbf  H_n$, is a square matrix with dimensions $n\times n$ consisting of elements of $\pm 1$.
It satisfies the property that the matrix product of $\mathbf  H_n$ and its transpose, $\mathbf  H_n^T$, results in a scalar multiple of the identity matrix, $\mathbf  I_n$, where $\mathbf  I_n$ is an $n\times n$ identity matrix and $n$ is the order of the Hadamard matrix.
%the distinct columns of H are pairwise orthogonal, as are the rows.
% \begin{equation}
% H_1=[1],  \quad \mathbf  H_2=\left[\begin{array}{rr}
% 1 & 1 \\
% 1 & -1\nonumber
% \end{array}\right]
% \end{equation}
\begin{equation}
\displaystyle \mathbf  H_n=\displaystyle \left[\begin{array}{rr}
\mathbf  H_{n-1} & \mathbf  H_{n-1}\\
\mathbf  H_{n-1} & -\mathbf  H_{n-1}
\end{array}\right]\nonumber, \quad n= 1, 2, 4, 16, 32, \ldots
\end{equation}
% The Sylvester construction of Hadamard matrices is a widely recognized mathematical technique with well-established foundations. This construction method facilitates the generation of Hadamard matrices with prescribed dimensions, including notable sizes such as $1, 2, 4, 8, 16, 32$, and so forth. The utilization of this approach leads to the formation of linear codes with inherent linear properties.

%The Hadamard transform computation of a linear vector, denoted as $\mathbf  H_n$ where $\mathbf  u$ is a linear vector of length $n$, is referred to as the computation of $\mathbf  r$.\\

Consider the received sequence $\mathbf  u = (u_0,u_1, \ldots, u_{2^m-1}) \in \mathbb{F}_2$, and let $\mathbf  c = (c_0,c_1, \ldots, c_{2^m-1})\in \mathbb{F}_2$ be a codeword. The bipolar representation of $\mathbf  u$ is denoted as $\mathbf  U \in \{-1,+1\}$ and is defined as $\mathbf  U= (-1)^{\mathbf  u}$. Similarly, the bipolar representation of $\mathbf  c$ is denoted as $\mathbf  C$ and defined as $\mathbf  C= (-1)^{\mathbf  c}$.
The decoding algorithm involves computing the correlation between $\mathbf  U$ and $\mathbf  C_i$, denoted as $\Delta_i$, for each of the $2^{m}$ codewords $\mathbf  C_i= (-1)^{\mathbf  c_i}$. The final step is to select the codeword for which $\Delta_i$ is the maximum. The simultaneous computation of all correlations can be depicted as a matrix representation. Denoting the column vector $C_i$ and constructing the matrix
$\mathbf  H=\left[\begin{array}{llll}\mathbf {C}_0 & \mathbf {C}_1 & \ldots & \mathbf {C}_{2^{m}-1}\end{array}\right]$, the computation of all correlations can be expressed as follows:
\begin{equation}
\mathbf {\Delta}=\mathbf {U} \mathbf  H,
\end{equation}
Where $\mathbf  H$ is a matrix of dimension $2^m$.
For the first subblock, which utilizes a $RM(1, 4)$ code with generator matrix $\mathbf  G^{(1)}$,  $\mathbf  H^{(1)}_{16}$ is employed. Similarly, for the second subblock, which employs a $RM(1, 5)$ code with generator matrix $\mathbf  G^{(2)}$, $\mathbf  H^{(2)}_{32}$ is utilized. Furthermore, detailed expositions concerning the algorithms employed in the decoding process of first-order RM codes through the utilization of the Hadamard transform can be readily found within the scholarly works authored by Moon \cite{Moon2005} and Wicker \cite{Wicker94}.

% The procedure of block-based RM decoding is therefore depicted Algorithm 1.
This decoding process  can be optimized through the utilization of a FHT which is applicable to Hadamard matrices produced through the Sylvester construction. This optimization is based on the fact that $\mathbf  H_{2^m}=\mathbf  H_2 \otimes \mathbf  H_{2^{m-1}}$, where the Kronecker product of matrices, denoted by $\otimes$, is applied. As a result, the matrix $\mathbf  H_{2^m}$ can be decomposed as stated in the theorem derived from linear algebraic principles\cite{Moon2005}.
 \begin{equation}
 \begin{aligned}
 & \mathbf  H_{2^m}=\mathbf  W_{2^m}^{(1)} \mathbf  W_{2^m}^{(2)} \cdots \mathbf  W_{2^m}^{(m)},
 \end{aligned}
 \end{equation}
 where $\mathbf  W_{2^m}^{(i)}=\mathbf  I_{2^{m-i}} \otimes \mathbf  H_2 \otimes \mathbf  I_{2^{i-1}}$, $\mathbf  I$ is an identity  matrix.
 \\
 Thus it comes,
 \begin{equation}
   \begin{aligned}
  \mathbf  H^{(1)}_{16}&=\mathbf  W_{16}^{(1)} \mathbf  W_{16}^{(2)} \mathbf  W_{16}^{(3)} \mathbf  W_{16}^{(4)}\\
 &=\left(\mathbf  I_{2^3} \otimes \mathbf  H_2 \otimes \mathbf  I_{2^0}\right)\left(\mathbf  I_{2^2} \otimes \mathbf  H_2 \otimes \mathbf  I_{2^1}\right)\\&\quad\left(\mathbf  I_{2^1} \otimes \mathbf  H_2 \otimes \mathbf  I_{2^2}\right)\left(\mathbf  I_{2^0} \otimes \mathbf  H_2 \otimes \mathbf  I_{2^3}\right).\quad \quad \quad \quad \quad \quad \quad \quad \quad \quad
 \end{aligned}
 \end{equation}
\begin{equation}
   \begin{aligned}
  \mathbf  H^{(2)}_{32}&=\mathbf  W_{32}^{(1)} \mathbf  W_{32}^{(2)} \mathbf  W_{32}^{(3)} \mathbf  W_{32}^{(4)} \mathbf  W_{32}^{(5)}\\
 &=\left(\mathbf  I_{2^4} \otimes \mathbf  H_2 \otimes \mathbf  I_{2^0}\right)\left(\mathbf  I_{2^3} \otimes \mathbf  H_2 \otimes \mathbf  I_{2^1}\right)\left(\mathbf  I_{2^2} \otimes \mathbf  H_2 \otimes \mathbf  I_{2^2}\right)\\&\quad\left(\mathbf  I_{2^1} \otimes \mathbf  H_2 \otimes \mathbf  I_{2^3}\right)\left(\mathbf  I_{2^0} \otimes \mathbf  H_2 \otimes \mathbf  I_{2^4}\right).
 \end{aligned}
 \end{equation}
Let's consider $
 \mathbf  U^{(1)} = [U_0, U_1,\ldots , U_{15}] \quad \text{and} \quad  \mathbf  U^{(2)} = [U_0, U_1,\ldots , U_{31}]$, the received sequences to be fed to the decoders.
 The corresponding Hadamard transforms can then be written as
 \begin{equation}
    \begin{aligned}
 \mathbf {\Delta}^{(1)}&=\mathbf {U}^{(1)} \mathbf  H^{(1)}_{16}=\mathbf {U}^{(1)}\left(\mathbf  W_{16}^{(1)} \mathbf  W_{16}^{(2)} \mathbf  W_{16}^{(3)} \mathbf  W_{16}^{(4)}\right).\\
  \mathbf {\Delta}^{(2)}&=\mathbf {U}^{(2)}  \mathbf  H^{(2)}_{32}=\mathbf {U}^{(2)}\left(\mathbf  W_{32}^{(1)} \mathbf  W_{32}^{(2)} \mathbf  W_{32}^{(3)} \mathbf  W_{32}^{(4)}\mathbf  W_{32}^{(5)}\right).
  \end{aligned}
 \end{equation}
 where
 \begin{equation}
   \begin{aligned}
 \mathbf  W_{16}^{(1)}=&\mathbf I_8 \otimes \mathbf  H_2, \quad   &\mathbf  W_{32}^{(1)}=&\mathbf I_{16} \otimes \mathbf  H_2, \\
 \mathbf  W_{16}^{(2)}=&\mathbf I_4 \otimes \mathbf  H_2 \otimes \mathbf I_2, \quad &\mathbf  W_{32}^{(2)}=&\mathbf I_8 \otimes \mathbf  H_2 \otimes \mathbf I_2 \\
 \mathbf  W_{16}^{(3)}=&\mathbf I_2 \otimes  \mathbf  H_2 \otimes \mathbf I_4, \quad  &\mathbf  W_{32}^{(3)}=&\mathbf I_4 \otimes  \mathbf  H_2 \otimes \mathbf I_4,\\
 \mathbf  W_{16}^{(4)}&=\mathbf H_2 \otimes \mathbf I_8, \quad &\mathbf  W_{32}^{(4)}=&\mathbf I_2\otimes \mathbf  H_2 \otimes \mathbf I_8\\
 &\quad &\mathbf  W_{32}^{(5)}&=\mathbf  H_2 \otimes \mathbf I_{16}.
 \end{aligned}\nonumber
 \end{equation}
%  \begin{figure}[!ht]
%         \centering
%       \includegraphics[width=\linewidth]{IEEE AFRICON 2023/figures/tikzfigure/algobb.pdf}
% \end{figure}

\begin{figure*}[!ht]
        \centering
        \includegraphics[width=0.9\linewidth]{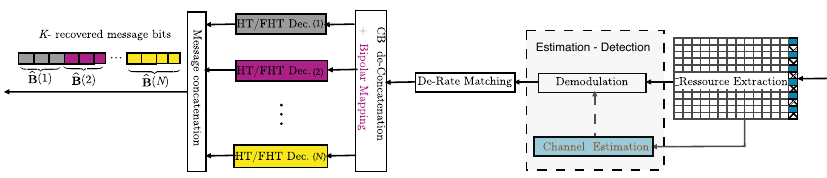}
        \caption{ Block-based Hadamard/Fast Hadamard Transform based decoding: Receiver end.}
       \label{fig:rmfirt_order_rx}
\end{figure*}

 The conventional computation of the Hadamard transform $\mathbf  H_{2^m}$ results in $2^m$ elements, each of which is obtained through $2^m$ addition/subtraction operations. This leads to a computational complexity of ${(2^m)}^2=\mathcal O\left({N^\prime}^2\right)$, which is equivalent to the complexity of a standard ML decoder that operates in a {\em quadratic order}. In contrast, the Fast Hadamard transform, which has $m$ stages, has a computational complexity of $m2^m=\mathcal O\left(N^\prime\log N^\prime\right)$ (i.e., {\em quasi-linear complexity})  due to its $2^m$ addition/subtraction operations per stage.
  The procedure of block-based RM decoding is therefore depicted in Figure \ref{fig:rmfirt_order_rx}.
\section{Numerical Results }
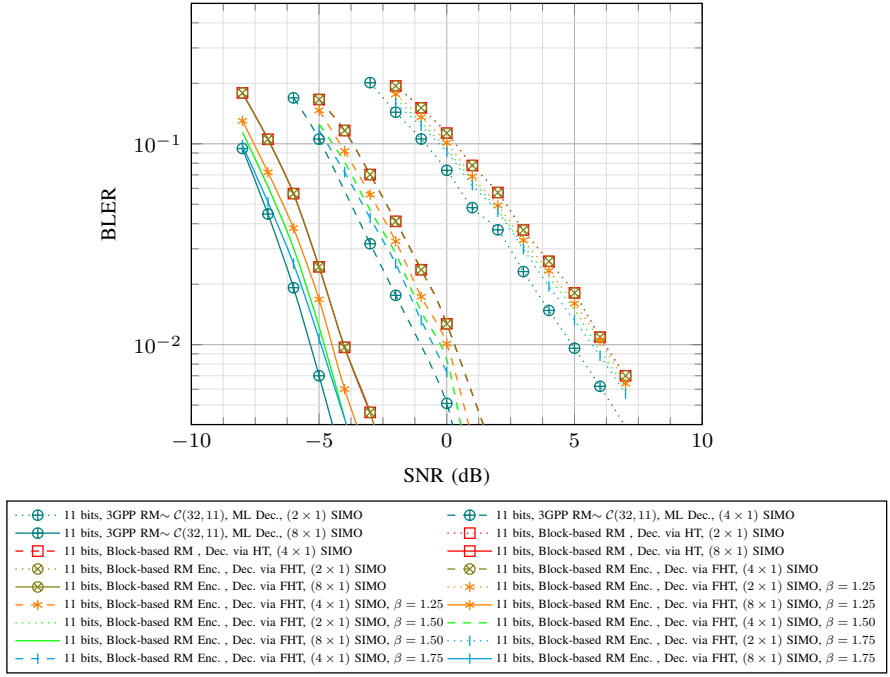
\begin{figure*}[!ht]
        \centering
        % \pgfplotsset{
%     tick label style={font=\footnotesize},
%     label style={font=\footnotesize},
%     legend style={nodes={scale=0.7, transform shape}, font=\tiny, legend cell align=left},
%     every axis/.append style={line width=0.5pt}
% }

%legend pos=outer north east
%legend pos=north west
%legend pos=north east
\pgfplotsset{
    tick label style={font=\footnotesize},
    label style={font=\footnotesize},
    legend style={nodes={scale=0.6, transform shape}, font=\footnotesize, legend cell align=left},
    every axis/.append style={line width=0.5pt}
}

%legend pos=outer north east
%legend pos=north west
%legend pos=north east

      \begin{tikzpicture}
              \begin{semilogyaxis}[xlabel=SNR (dB),
            ,ylabel=BLER,
            %legend pos=south west,
            legend columns=2,
            legend style={at={(0.5,-.18)},anchor=north},
             grid=both,
            minor tick num=3,
             xmin=-10,xmax=10, ymin=0.005,ymax=0.5]

\addplot+[smooth,teal, dotted, mark options={scale=1, fill=teal, solid},mark=oplus] coordinates {
(-3,	0.2016)
(-2,	0.1436)
(-1,	0.1057)
(0,	0.0738)
(1,	0.0481)
(2,	0.0373)
(3,	0.0231)
(4,	0.0148)
(5,	0.0096)
(6,	0.0062)
(7,	0.0039)
(8,	0.0022)
(9,	0.002)
(10,	0.0015)
            };
\addlegendentry{11 bits, 3GPP RM$\sim \mathcal C(32,11)$,  ML Dec., $(2\times1)$ SIMO}

\addplot+[smooth,teal, dashed, mark options={scale=1, fill=teal, solid},mark=oplus] coordinates {
(-6,	0.1693)
(-5,	0.1055)
(-4,	0.063)
(-3,	0.0318)
(-2,	0.0176)
(-1,	0.0089)
(0,	0.0051)
(1,	0.0012)
(2,	0.0009)
(3,	0.0004)
(4,	0.0001)
(5,	0.0001)
};
\addlegendentry{11 bits, 3GPP RM$\sim \mathcal C(32,11)$,  ML Dec., $(4\times1)$ SIMO}

\addplot+[smooth,teal, solid, mark options={scale=1, fill=teal, solid},mark=oplus] coordinates {
%(-9,	0.175)
(-8,	0.0947)
(-7,	0.0447)
(-6,	0.0192)
(-5,	0.007)
(-4,	0.0023)
(-3,	0.0005)
(-2,	0.0002)
(-1,	0.0001)
};
\addlegendentry{11 bits, 3GPP RM$\sim \mathcal C(32,11)$,  ML Dec., $(8\times1)$ SIMO}

%
% \addplot+[smooth,teal, dashed, mark options={scale=1, fill=teal, solid},mark=oplus] coordinates {
% (-5,	0.127)
% (-4,	0.065)
% (-3,	0.047)
% (-2,	0.022)
% (-1,	0.011)
% (0	,0.006)
% };
% \addlegendentry{11 bits, 3GPP RM$\sim \mathcal C(32,11)$,  ML Dec., $(4\times1)$ SIMO}
%
%
% \addplot+[smooth,teal, solid, mark options={scale=1, fill=teal, solid},mark=oplus] coordinates {
% %(-9,	0.175)
% (-8,	0.125)
% (-7,	0.057)
% (-6,	0.019)
% (-5,	0.01)
% (-4,	0.003)
% };
% \addlegendentry{11 bits, 3GPP RM$\sim \mathcal C(32,11)$,  ML Dec., $(8\times1)$ SIMO}

%%%%%%%%%%%%%%%%%%%%%%%%%%%%%%%%%%%%%%%%%%%%%%%%
\addplot+[smooth, dotted, red, mark options={ scale=1, fill=red, solid},mark=square] coordinates {
(-2, 	0.1941)
(-1, 	0.151)
(0	, 0.113)
(1	, 0.078)
(2	, 0.0572)
(3	, 0.0372)
(4	, 0.026)
(5	, 0.0181)
(6	, 0.0109)
(7	, 0.007)
(8	, 0.0041)
(9	, 0.0036)
(10, 	0.0022)
};
\addlegendentry{11 bits, Block-based RM , Dec. via HT, $(2\times1)$ SIMO}

\addplot+[smooth, dashed, red, mark options={ scale=1, fill=red, solid},mark=square] coordinates {
(-5, 	0.1662)
(-4, 	0.1164)
(-3, 	0.0703)
(-2, 	0.0411)
(-1, 	0.0236)
(0	, 0.0127)
(1	, 0.0048)
(2	, 0.0038)
(3	, 0.0011)
(4	, 0.0005)
(5	, 0.0002)
(6	, 0.0001)
(7	, 0.0001)
};
\addlegendentry{11 bits, Block-based RM , Dec. via HT, $(4\times1)$ SIMO}

\addplot+[smooth, solid, red, mark options={ scale=1, fill=red, solid},mark=square] coordinates {
(-8,	0.1792)
(-7,	0.1053)
(-6,	0.0565)
(-5,	0.0244)
(-4,	0.0097)
(-3,	0.0046)
(-2,	0.0016)
(-1,	0.0007)
(0	,0.0001)
};
\addlegendentry{11 bits, Block-based RM , Dec. via HT, $(8\times1)$ SIMO}

%%%%%%%%%%%%%%%%%%%%%%%%%%%%%

\addplot+[smooth, dotted, violet, mark options={ scale=1, fill=violet, solid},mark=otimes] coordinates {
(-2, 	0.1941)
(-1, 	0.151)
(0	, 0.113)
(1	, 0.078)
(2	, 0.0572)
(3	, 0.0372)
(4	, 0.026)
(5	, 0.0181)
(6	, 0.0109)
(7	, 0.007)
(8	, 0.0041)
(9	, 0.0036)
(10, 	0.0022)
};
\addlegendentry{11 bits, Block-based RM Enc. , Dec. via FHT, $(2\times1)$ SIMO}

\addplot+[smooth, dashed, violet, mark options={ scale=1, fill=violet, solid },mark=otimes] coordinates {
(-5, 	0.1662)
(-4, 	0.1164)
(-3, 	0.0703)
(-2, 	0.0411)
(-1, 	0.0236)
(0	, 0.0127)
(1	, 0.0048)
(2	, 0.0038)
(3	, 0.0011)
(4	, 0.0005)
(5	, 0.0002)
(6	, 0.0001)
(7	, 0.0001)
};
\addlegendentry{11 bits, Block-based RM Enc. , Dec. via FHT, $(4\times1)$ SIMO}

\addplot+[smooth, solid, violet, mark options={ scale=1, fill=violet},mark=otimes] coordinates {
(-8,	0.1792)
(-7,	0.1053)
(-6,	0.0565)
(-5,	0.0244)
(-4,	0.0097)
(-3,	0.0046)
(-2,	0.0016)
(-1,	0.0007)
(0	,0.0001)
};
\addlegendentry{11 bits, Block-based RM Enc. , Dec. via FHT, $(8\times1)$ SIMO}

%%%%%%%%%%%%%%%%%%%%%%%%%%%%%%%%%%%%%%%%%%%%%%%%%%%%%
\addplot+[smooth,orange, dotted, mark options={scale=1, fill=orange, solid},mark=asterisk] coordinates {
(-2, 	0.1763)
(-1, 	0.1355)
(0	, 0.1011)
(1	, 0.0687)
(2	, 0.0493)
(3	, 0.0331)
(4	, 0.0233)
(5	, 0.016)
(6	, 0.0105)
(7	, 0.0064)
(8	, 0.0038)
(9	, 0.0026)
(10, 	0.0019)
};
\addlegendentry{11 bits, Block-based RM Enc. , Dec. via FHT, $(2\times1)$ SIMO, $\beta = 1.25$}

\addplot+[smooth,orange, dashed, mark options={scale=1, fill=orange, solid},mark=asterisk] coordinates {
(-5, 	0.1467)
(-4, 	0.092)
(-3, 	0.0558)
(-2, 	0.0327)
(-1, 	0.0173)
(0	, 0.0101)
(1	, 0.0034)
(2	, 0.0031)
(3	, 0.0007)
(4	, 0.0004)
(5	, 0.0002)
(6	, 0)
(7	, 0.0001)
(8	, 0)
(9	, 0)
(10, 	0)
};
\addlegendentry{11 bits, Block-based RM Enc. , Dec. via FHT, $(4\times1)$ SIMO, $\beta = 1.25$}

\addplot+[smooth,orange, solid, mark options={scale=1, fill=orange, solid},mark=asterisk] coordinates {
(-8, 	0.1301)
(-7, 	0.0721)
(-6, 	0.038)
(-5, 	0.0168)
(-4, 	0.006)
(-3, 	0.0026)
(-2, 	0.0008)
(-1, 	0.0003)
(0	, 0.0001)

};
\addlegendentry{11 bits, Block-based RM Enc. , Dec. via FHT, $(8\times1)$ SIMO, $\beta = 1.25$}

%%%%%%%%%%%%%%%%%%%%%%%%%%%%%%%%%%%%
\addplot+[smooth,green, dotted, mark options={scale=1, fill=green, solid},mark=]coordinates {
(-2, 	0.1673)
(-1, 	0.1277)
(0	, 0.094)
(1	, 0.0657)
(2	, 0.0468)
(3	, 0.031)
(4	, 0.0216)
(5	, 0.0149)
(6	, 0.0093)
(7	, 0.0062)
(8	, 0.0034)
(9	, 0.0019)
(10, 	0.0017)
};
\addlegendentry{11 bits, Block-based RM Enc. , Dec. via FHT, $(2\times1)$ SIMO, $\beta = 1.50$}

\addplot+[smooth,green, dashed, mark options={scale=1, fill=green, solid},mark=]coordinates {
%(-6, 	0.1851)
(-5, 	0.1255)
(-4, 	0.0812)
(-3, 	0.0468)
(-2, 	0.0284)
(-1, 	0.0146)
(0	, 0.0084)
(1	, 0.0023)
(2	, 0.0022)
(3	, 0.0008)
(4	, 0.0004)
(5	, 0.0001)
(6	, 0)
(7,	0.0001)
};
\addlegendentry{11 bits, Block-based RM Enc. , Dec. via FHT, $(4\times1)$ SIMO, $\beta = 1.50$}
\addplot+[smooth,green, solid, mark options={scale=1, fill=green, solid},mark=]coordinates {
(-8, 	0.1142)
(-7, 	0.0613)
(-6, 	0.0303)
(-5, 	0.0123)
(-4, 	0.0043)
(-3, 	0.002)
(-2, 	0.0008)
(-1, 	0.0003)
};
\addlegendentry{11 bits, Block-based RM Enc. , Dec. via FHT, $(8\times1)$ SIMO, $\beta = 1.50$}

%%%%%%%%%%%%%%%%%%%%%%%%%%%%%%%%%%%%

\addplot+[smooth,cyan, dotted, mark options={scale=1, fill=cyan, solid},mark=|]coordinates {
%-7	0.2666)
(-2, 	0.1589)
(-1, 	0.1226)
(0	, 0.0912)
(1	, 0.0622)
(2	, 0.0459)
(3	, 0.0299)
(4	, 0.0195)
(5	, 0.0132)
(6	, 0.0088)
(7	, 0.0057)
(8	, 0.0033)
(9	, 0.0017)
};
\addlegendentry{11 bits, Block-based RM Enc. , Dec. via FHT, $(2\times1)$ SIMO, $\beta = 1.75$}

\addplot+[smooth,cyan, dashed, mark options={scale=1, fill=cyan, solid},mark=|]coordinates {
%-7	0.2666)
(-5, 	0.1165)
(-4, 	0.0722)
(-3, 	0.0427)
(-2, 	0.0253)
(-1, 	0.0132)
(0	, 0.0073)
(1	, 0.0023)
(2	, 0.0021)
(3	, 0.0009)
(4	, 0.0003)
(5	, 0.0001)
};
\addlegendentry{11 bits, Block-based RM Enc. , Dec. via FHT, $(4\times1)$ SIMO, $\beta = 1.75$}

\addplot+[smooth,cyan, solid, mark options={scale=1, fill=cyan, solid},mark=|]coordinates {
%-7	0.2666)
%(-9, 	0.1547)
(-8, 	0.0984)
(-7	, 0.0512)
(-6,	0.0253)
(-5,	0.0107)
(-4,	0.0028)
(-3,	0.0017)
(-2,	0.0005)
(-1,	0.0002)
};
\addlegendentry{11 bits, Block-based RM Enc. , Dec. via FHT, $(8\times1)$ SIMO, $\beta = 1.75$}

%%%%%%%
        \end{semilogyaxis}
    \end{tikzpicture}
        \caption{Block Error Rate, 11 bits, 2 PRB(16 REs=data, 8 REs=DMRS), Block-based  decoding via  HT \& FHT based decoders vs ML decoder, Adaptative DMRS/data Power  Adjustment bia $\beta$, ($\{8, 4, 2\} \times 1$) SIMO, Rayleigh fading Channel, Unknown {\em Channel State Information}.}
        \label{fig:adaptative_power_ajustement11bits}
\end{figure*}

For illustration purposes, we focuse on PUCCH2-based short block lengths, which consist of 11 bits. PUCCH2 is configurable in terms of the resource usage, but we consider the simplest comprising of $2$ groups of 12 dimensions or {\em resource elements}, so-called  PRBs, making $24$ dimensions which consist of $16$ for data components, and $8$ for the so-called  DMRS, which are known symbols used for channel estimation and tracking.
The following simulation results were performed under the assumption of Rayleigh flat fading channel, utilizing 2, 4 and 8 antenna configurations. The antenna ports were subjected to independent and identically distributed realizations, with no correlation modeling applied. Figure~\ref{fig:adaptative_power_ajustement11bits}, demonstrate the merits of the proposed block-based Reed-Muller (RM) decoding approach employing the HT/FHT. These results are contrasted with the conventional RM decoding technique recommended by 3GPP, employing a {\em Maximum Likelihood} (ML) decoder, within the Rayleigh fading channel scenario.

% \begin{figure}[!ht]
%         \centering
%           %\includegraphics[width=0.75\linewidth]{figures/fig_pdf/fht_optim.pdf}
%         \input{figures/tikz/ray/comp_optim.tex}
%         \caption{Block Error Rate, 11 bits, 2 PRB(16 REs=data, 8 REs=DMRS), Block-based  decoding via  HT \& FHT based decoders vs ML decoder, Adaptative DMRS/data Power  Adjustment,  Unknown {\em Channel State Information} (CSI), nRx = 2,4, 8, Rayleigh fading Channel. }
%         \label{fig:adaptative_power_ajustement11bits}
% \end{figure}
The results presented suggest that the ML decoder exhibits improved performance, however, this advantage is accompanied by a significant increase in computational complexity, even when utilizing the block decoding principle.
In practical implementation, each received code block ($\mathcal{C}(11,48)=[ \mathcal{C}_1(5,16), \mathcal{C}_2(6,32)$)]) is processed through a corresponding  decoder, followed by concatenation at the decoder output stage. It is important to acknowledge that the efficacy of block decoding through both HT and FHT is comparable, exhibiting a marginal deviation of $1.3$ dB from the ML receiver at a BLER threshold of $1\%$ when nRx=$4$.
Moreover, some of the performance gap with the  ML receiver can be bridged or at least reduced. 

\begin{figure}[!ht]
        \centering
       % \pgfplotsset{
%     tick label style={font=\footnotesize},
%     label style={font=\footnotesize},
%     legend style={nodes={scale=0.7, transform shape}, font=\tiny, legend cell align=left},
%     every axis/.append style={line width=0.5pt}
% }

% legend pos=outer north east
% legend pos=north west
% legend pos=north east
\pgfplotsset{
    tick label style={font=\footnotesize},
    label style={font=\footnotesize},
    legend style={nodes={scale=0.7, transform shape}, font=\tiny, legend cell align=left},
    every axis/.append style={line width=0.5pt}
}

%legend pos=outer north east
%legend pos=north west
%legend pos=north east

      \begin{tikzpicture}
              \begin{semilogyaxis}[xlabel=SNR (dB),
            ,ylabel=BLER,
            legend pos=north east,
            minor tick num=3,
            grid=both, xmin=-2,xmax=10, ymin=0.001,ymax=0.3]

%%%%%%%%%%%%%%%%%%%%%%%%%%%%%%%%%%%%%%
% \pgfplotsset{
%     tick label style={font=\footnotesize},
%     label style={font=\footnotesize},
%     legend style={nodes={scale=0.50, transform shape}, font=\tiny, legend cell align=left},
%     every axis/.append style={line width=0.5pt}
% }
%
% %legend pos=outer north east
% %legend pos=north west
% %legend pos=north east
%
%       \begin{tikzpicture}
%               \begin{semilogyaxis}[xlabel=SNR (dB),
%             ,ylabel=BLER,
%             %legend pos=south west,
%             legend columns=2,
%             legend style={at={(0.5,1.02)},anchor=south},
%             grid=both, xmin=-2,xmax=12, ymin=0.002,ymax=0.3]

%%%%%%%%%%%%%%%%%%%%%%%%%%%%%%%%%

% \addplot+[smooth,teal, dotted, mark options={scale=1, fill=teal, solid},mark=oplus] coordinates {
% (0	,0.147275405007364)
% %(1	,0.0922509225092251)
% (2	,0.0769230769230769)
% (3	,0.0570125427594071)
% (4	,0.0432525951557093)
% (5	,0.0264900662251656)
% (6	,0.0195274360476469)
% (7	,0.0133547008547009)
% (8	,0.010418837257762)
% %(9	,0.0057)
% % (10,	0.0045)
% (11,	0.0025)
% (12,	0.0016)
%             };
% \addlegendentry{11 bits, 3GPP RM$\sim \mathcal C(32,11)$,  ML Dec., nTx=2, $(8\times2)$ SIMO}

\addplot+[smooth,teal, dashed, mark options={scale=1, fill=teal, solid},mark=oplus] coordinates {
(-1, 	0.127226463104326)
(0	, 0.0731528895391368)
(1	, 0.0548847420417124)
(2	, 0.0370233246945576)
(3	, 0.0230414746543779)
(4	, 0.0130276185513288)
(5	, 0.0078)
(6	, 0.004)
(7	, 0.0019)
(8	, 0.001)
};
\addlegendentry{11 bits, 3GPP RM$\sim \mathcal C(32,11)$,  ML Dec., $(4\times4)$ MIMO}

%
% \addplot+[smooth,teal, solid, mark options={scale=1, fill=teal, solid},mark=oplus] coordinates {
% %(-9,	0.175)
% (-8,	0.0947)
% (-7,	0.0447)
% (-6,	0.0192)
% (-5,	0.007)
% (-4,	0.0023)
% (-3,	0.0005)
% (-2,	0.0002)
% (-1,	0.0001)
% };
% \addlegendentry{11 bits, 3GPP RM$\sim \mathcal C(32,11)$,  ML Dec., $(8\times1)$ SIMO}

% %%%%%%%%%%%%%%%%%%%%%%%%%%%%%%%%%%%%%%%%%%%%%%%%
% \addplot+[smooth, dotted, red, mark options={ scale=1, fill=red, solid},mark=square] coordinates {
% (1, 0.143266475644699)
% %(2, 0.0783699059561129)
% (3, 0.0716332378223496)
% % (4, 0.0472813238770686)
% % (5, 0.0336813742000674)
% % (6,	0.0231481481481481)
% (7	,0.0193610842207164)
% (8	,0.0118077695123391)
% (9	,0.0082)
% (10,	0.0049)
% (11,	0.0034)
% (12,	0.0017)
% };
% \addlegendentry{11 bits, Block-based RM , Dec. via HT, nTx=2 $(8\times2)$ SIMO}

\addplot+[smooth, dashed, red, mark options={ scale=1, fill=red, solid},mark=square] coordinates {
(-1, 	0.157977883096367)
%(0	, 0.0860585197934595)
%(1	, 0.0808407437348424)
%(2	, 0.0595238095238095)
%(3	, 0.0269832703723691)
(4	, 0.0226244343891403)
(5	, 0.0124875124875125)
(6	, 0.0071)
(7	, 0.0033)
(8	, 0.0019)
(9	, 0.0011)
};
\addlegendentry{11 bits, Block-based RM , Dec. via HT, $(4\times4)$ MIMO}
%
% \addplot+[smooth, solid, red, mark options={ scale=1, fill=red, solid},mark=square] coordinates {
% (-8,	0.1792)
% (-7,	0.1053)
% (-6,	0.0565)
% (-5,	0.0244)
% (-4,	0.0097)
% (-3,	0.0046)
% (-2,	0.0016)
% (-1,	0.0007)
% (0	,0.0001)
% };
% \addlegendentry{11 bits, Block-based RM , Dec. via HT, $(8\times1)$ SIMO}
%
% %%%%%%%%%%%%%%%%%%%%%%%%%%%%%

% \addplot+[smooth, dotted, violet, mark options={ scale=1, fill=violet, solid},mark=otimes] coordinates {
% (1, 0.143266475644699)
% %(2, 0.0783699059561129)
% (3, 0.0716332378223496)
% % (4, 0.0472813238770686)
% % (5, 0.0336813742000674)
% % (6,	0.0231481481481481)
% (7	,0.0193610842207164)
% (8	,0.0118077695123391)
% (9	,0.0082)
% (10,	0.0049)
% (11,	0.0034)
% (12,	0.0017)
% };
% \addlegendentry{11 bits, Block-based RM Enc. , Dec. via FHT, nTx=2, $(8\times2)$ SIMO}

\addplot+[smooth, dashed, violet, mark options={ scale=1, fill=violet, solid },mark=otimes] coordinates {
(-1, 	0.157977883096367)
% (0	, 0.0860585197934595)
% (1	, 0.0808407437348424)
% (2	, 0.0595238095238095)
%(3	, 0.0269832703723691)
(4	, 0.0226244343891403)
(5	, 0.0124875124875125)
(6	, 0.0071)
(7	, 0.0033)
(8	, 0.0019)
(9	, 0.0011)
};
\addlegendentry{11 bits, Block-based RM Enc. , Dec. via FHT, $(4\times4)$ MIMO}
%
% \addplot+[smooth, solid, violet, mark options={ scale=1, fill=violet},mark=otimes] coordinates {
% (-8,	0.1792)
% (-7,	0.1053)
% (-6,	0.0565)
% (-5,	0.0244)
% (-4,	0.0097)
% (-3,	0.0046)
% (-2,	0.0016)
% (-1,	0.0007)
% (0	,0.0001)
% };
% \addlegendentry{11 bits, Block-based RM Enc. , Dec. via FHT, $(8\times1)$ SIMO}
%
%
% %%%%%%%%%%%%%%%%%%%%%%%%%%%%%%%%%%%%%%%%%%%%%%%%%%%%%
% \addplot+[smooth,orange, dotted, mark options={scale=1, fill=orange, solid},mark=asterisk] coordinates {
% (0	,0.141643059490085)
% (1	,0.106723585912487)
% (2	,0.0699300699300699)
% (3	,0.0530222693531283)
% (4	,0.0403063280935107)
% (5	,0.0293513354857646)
% (6	,0.0199044585987261)
% (7	,0.0136724090784796)
% (8	,0.0110497237569061)
% (9	,0.0063)
% (10,	0.004)
% (11,	0.0028)
% (12,	0.0017)
% };
% \addlegendentry{11 bits, Block-based RM Enc. , Dec. via FHT, $(4\times4)$ MIMO, $\beta = 1.25$}

\addplot+[smooth,orange, dashed, mark options={scale=1, fill=orange, solid},mark=asterisk] coordinates {
(-1, 	0.102354145342886)
(0	, 0.0794912559618442)
(1	, 0.0558347292015634)
(2	, 0.0323415265200517)
(3	, 0.0212089077412513)
(4	, 0.0135263086703639)
(5	, 0.0073)
(6	, 0.0043)
(7	, 0.0024)
(8	, 0.0012)
};
\addlegendentry{11 bits, Block-based RM Enc. , Dec. via FHT, $(4\times4)$ MIMO, $\beta = 1.25$}

% \addplot+[smooth,orange, solid, mark options={scale=1, fill=orange, solid},mark=asterisk] coordinates {
% (-8, 	0.1301)
% (-7, 	0.0721)
% (-6, 	0.038)
% (-5, 	0.0168)
% (-4, 	0.006)
% (-3, 	0.0026)
% (-2, 	0.0008)
% (-1, 	0.0003)
% (0	, 0.0001)
%
% };
% \addlegendentry{11 bits, Block-based RM Enc. , Dec. via FHT, $(8\times1)$ SIMO, $\beta = 1.25$}
%
% %%%%%%%%%%%%%%%%%%%%%%%%%%%%%%%%%%%%
% \addplot+[smooth,green, dotted, mark options={scale=1, fill=green, solid},mark=]coordinates {
% (1	,0.102669404517454)
% (2	,0.0660501981505945)
% (3	,0.050251256281407)
% (4	,0.0383582662063675)
% (5	,0.03125)
% (6	,0.0192826841496336)
% (7	,0.0129954515919428)
% (8	,0.0089)
% (9	,0.0066)
% (10,	0.0042)
% (11,	0.0018)
% (12,	0.0022)
%
% };
% \addlegendentry{11 bits, Block-based RM Enc. , Dec. via FHT, $(4\times4)$ MIMO, $\beta = 1.50$}

\addplot+[smooth,green, dashed, mark options={scale=1, fill=green, solid},mark=]coordinates {
%(-6, 	0.1851)
%(-2, 	0.132802124833997)
(-1, 	0.0809061488673139)
(0 ,0.0639386189258312)
(1	, 0.043122035360069)
(2	, 0.0243072435585805)
%(3	, 0.0208289939595918)
(4	, 0.0081)
(5	, 0.0051)
(6	, 0.0031)
(7	, 0.0017)
};
 \addlegendentry{11 bits, Block-based RM Enc. , Dec. via FHT, $(4\times4)$ MIMO, $\beta = 1.50$}
% \addplot+[smooth,green, solid, mark options={scale=1, fill=green, solid},mark=]coordinates {
% (-8, 	0.1142)
% (-7, 	0.0613)
% (-6, 	0.0303)
% (-5, 	0.0123)
% (-4, 	0.0043)
% (-3, 	0.002)
% (-2, 	0.0008)
% (-1, 	0.0003)
% };
% \addlegendentry{11 bits, Block-based RM Enc. , Dec. via FHT, $(8\times1)$ SIMO, $\beta = 1.50$}

%%%%%%%%%%%%%%%%%%%%%%%%%%%%%%%%%%%%
%
% \addplot+[smooth,cyan, dotted, mark options={scale=1, fill=cyan, solid},mark=|]coordinates {
% %-7	0.2666)
% (1	,0.108459869848156)
% (2	,0.06426735218509)
% (3	,0.0496524329692155)
% (4	,0.0358680057388809)
% (5	,0.0312304809494066)
% (6	,0.0207125103562552)
% (7	,0.0134916351861846)
% (8	,0.0098)
% (9	,0.0066)
% (10,	0.0037)
% (11,	0.0018)
% (12,	0.0019)
% };
% \addlegendentry{11 bits, Block-based RM Enc. , Dec. via FHT, $(4\times4)$ MIMO, $\beta = 1.75$}

\addplot+[smooth,cyan, dashed, mark options={scale=1, fill=cyan, solid},mark=|]coordinates {
%-7	0.2666)
%(-2, 	0.121359223300971)
(-1, 	0.0711237553342817)
(0	, 0.0530222693531283)
(1	, 0.0341296928327645)
(2	, 0.0228990153423403)
%(3	, 0.01669727834363)
(4	, 0.007)
(5	, 0.0039)
(6	, 0.0026)
(7	, 0.0013)
};
\addlegendentry{11 bits, Block-based RM Enc. , Dec. via FHT, $(4\times4)$ MIMO, $\beta = 1.75$}

% \addplot+[smooth,cyan, solid, mark options={scale=1, fill=cyan, solid},mark=|]coordinates {
% %-7	0.2666)
% %(-9, 	0.1547)
% (-8, 	0.0984)
% (-7	, 0.0512)
% (-6,	0.0253)
% (-5,	0.0107)
% (-4,	0.0028)
% (-3,	0.0017)
% (-2,	0.0005)
% (-1,	0.0002)
% };
% \addlegendentry{11 bits, Block-based RM Enc. , Dec. via FHT, $(8\times1)$ SIMO, $\beta = 1.75$}

%%%%%%%
        \end{semilogyaxis}
    \end{tikzpicture}
        \caption{Block Error Rate, 11 bits, 2 PRB(16 REs=data, 8 REs=DMRS), Block-based  decoding via  HT \& FHT based decoders vs ML decoder, Adaptative DMRS/data Power  Adjustment,  Unknown {\em Channel State Information} (CSI), ( $4\times4$) MIMO, Rayleigh fading Channel. }
        \label{fig:sdt_deco_11_ray-MIMO}
\end{figure}
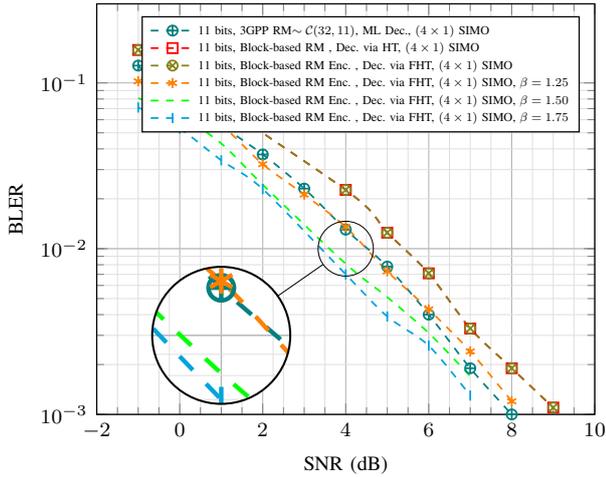

In instances where we transmit reference and data symbols jointly in common OFDM symbols, as is the case for PUCCH or PUSCH   or in some downlink control channels, there is the possibility to optimize the data-reference power ratio. This is possible for both downlink and uplink transmission without incurring a penalty in terms of peak power increase.\\
Therefore, the performance disparity between the ML receiver and the FHT-based receiver can be mitigated/reduced by reconceiving the system as follows  $\mathbf {y}=\left(\mathbf {x}^{(d)}+ \beta\mathbf {x}^{(p)}\right)\cdot\mathbf {h}+\mathbf {z}$.
This adaptive power adjustment procedure is contingent on the values of $\beta$. The DMRS power is to be slightly increased in a judicious manner since $\beta$ must be perfectly calibrated to ensure compliance with potential radio frequency constraints.
Figure~\ref{fig:adaptative_power_ajustement11bits} illustrates the observable performance enhancement achieved by varying $\beta$ values, specifically $\beta=\{ 1.25, 1.5, 1.75\}$. The most significant performance gain occurs when $\beta = 1.75$.
Through simulation, it has been demonstrated that when the number of receive antennas is 4 and $\beta = 1.75$ is selected, the performance of the FHT-based decoder approaches that of the ML decoder at a BLER threshold of $1\%$, corresponding to an additional gain of $1$ dB.
Furthermore, it is important to note that the adaptive DMRS/data power adjustment process yields greater benefits with an increased number of receiving antennas. Consequently, the disparity between the performance of the ML receiver and the FHT-based receiver can be minimized.\\
Likewise, the simulation outcomes pertaining to a fading channel at the link level, incorporating MIMO, are depicted in Figure~\ref{fig:sdt_deco_11_ray-MIMO}. The investigation encompasses a MIMO system operating with spatial multiplexing, specifically a $(4\times4)$ configuration.
We observe that there exists a performance disparity of $1$ dB between the ML receiver and the FHT-based receiver at a BLER of $1\%$.
The adaptive adjustment process allows an additional gain of approximately $1$ dB by selecting $\beta=1.50$, thus bridging the gap between the ML and FHT receivers, and $2$ dB by selecting $\beta=1.75$, outperforming the ML receiver by $1$ dB.\\
Overall, the implications of performing an adaptive DMRS/data power adjustment within the 3GPP standard are significant in term of performance improvement. Specifically, the  {\em User equipment} (UE) could vary  the power allocation between the DMRS and data. This flexibility in  power allocation is transparent to the receiver.\\

%%%%%%%%%%complexity analysis
Furthermore, in terms of computational efficiency, block decoding using the Fast Hadamard transform is more advantageous as it offers a faster convergence time, which we will endeavour to demonstrate hereafter.

\begin{figure}[!ht]
        \centering
        \pgfplotsset{
            tick label style={font=\footnotesize},
            label style={font=\footnotesize},
            legend style={nodes={scale=0.75, transform shape}, font=\tiny, legend cell align=left},
            every axis/.append style={line width=0.5pt}
        }
\begin{tikzpicture}

\begin{axis} [ybar = .4cm,
    bar width = 15pt,
    grid=both,
    minor tick num=3,
    %legend style={at={(0.5,-0.15)},
    %   anchor=north,legend columns=-1},
  % ybar=5pt,%  configures ‘bar shift’
    ymin = 0,
    ymax =2048,
    xmin = 0,
    xmax =1.75,
    enlarge x limits = {value = .25, upper},
    ylabel={ Number of operations ($\mathcal N$)},
    xlabel={Decoders},
    nodes near coords
]

\addplot [draw=none, fill=teal!90] coordinates {
  (1,2048)
  %(Oct-11, 72.7961)
 };
\addlegendentry{$\sqrt{\text{ML DEC.,} \mathcal O\left({N^\prime}^2\right)}$}

 \addplot [draw=none, fill=magenta!90] coordinates {
   (1,64)
  };
 \addlegendentry{FHT-DEC. for Block 1,  $\mathcal O\left({N_1^\prime}\log {N_1^\prime}\right)$}

 \addplot [draw=none, fill=gray!90] coordinates {
   (1,160)
  };
 \addlegendentry{FHT-DEC. for Block 2,  $\mathcal O\left({N_2^\prime}\log {N_2^\prime}\right)$}

  \addplot [draw=none, fill=orange!90] coordinates {
    (1,224)
   };
 \addlegendentry{FHT-DEC. (Seq. processing),  $\mathcal O\left({N_1^\prime}\log {N_1^\prime}\right) + \mathcal O\left( {N_2^\prime}\log {N_2^\prime}\right)$}

 \addplot [draw=none, fill=cyan!90] coordinates {
   (1,160)
  };

 \addlegendentry{FHT-DEC.(Par. Processing), $\mathcal O\left(\max\left({N_1^\prime}\log {N_1^\prime}, {N_2^\prime}\log {N_2^\prime}\right)\right)$}

\end{axis}

\end{tikzpicture}
\caption{Comparative analysis, showing the number of $\mathcal N$ operations for the ML decoder compared  to  block-based FHT-decoders featuring sequential and parallel processing.}
\label{fig:sdt_deco_11_compexity}
\end{figure}
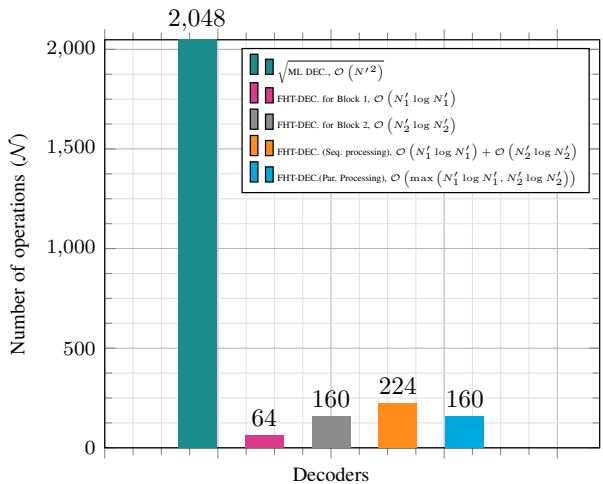

The graph in Figure~\ref{fig:sdt_deco_11_compexity} portrays the number of operations required, denoted as $\mathcal N$, in relation to the input size, denoted as $N^\prime=2^K$, for both the ML decoder and the FHT-based decoders. Notwithstanding the computational time complexity involved can be assessed through comprehensive numerical simulations, we approach this inquiry within a relatively simplified analytical or theoretical framework.

This method seeks to encapsulate the quantum of steps or operations essential for a decoder to achieve convergence and successfully retrieve transmitted messages. To aid comprehension of the graph, we introduce the notations ${N_1}^\prime = 2^4=16$ and ${N_2^\prime} = 2^5=32$, representing the number of candidate data words that can be retrieved for the FHT-based decoder 1 and FHT-based decoder 2. These notations correspond to $\mathcal N_1^\prime = ({N_1^\prime}\log {N_1^\prime})$ and $\mathcal N_2^\prime = {N_2^\prime}\log {N_2^\prime}$ operations. As for the ML decoder, which serves to recover encoded codewords using the 3GPP RM code family, the case is such that $N^\prime = 2^{11}=2048$,
 resulting in $\mathcal N^{(ML)} = {N^\prime}^2$ operations.

 Lastly, focusing on the block-based FHT decoder and initially assuming that decoders are executed in parallel and to some extent independently from each other, as illustrated in Figure \ref{fig:rmfirt_order_rx}, this strategy is employed to leverage the merit of parallelism in processing, thereby leading to a diminished inherent complexity.
For the block-based FHT with ${N^\prime} = {N_1^\prime} + {N_2^\prime}$, the overall decoding complexity is encapsulated as $\mathcal O\left(\max\left({N_1^\prime}\log {N_1^\prime}, {N_2^\prime}\log {N_2^\prime}\right)\right)$, equivalent to $\mathcal N^{(FHT)} = ({N_2^\prime}\log {N_2^\prime})$, representing the decoder associated with the longest codeword.

% \begin{figure}[!ht]
% \centering
% \includegraphics[width=0.9\linewidth]{figures/tikzfigure/6.pdf}
% \caption{Comparative analysis, showing the number of $\mathcal N$ operations for the ML decoder compared  to  block-based FHT-decoders featuring sequential and parallel processing.}
% \label{fig:sdt_deco_11_compexity}
% \end{figure}
% %185.82
The graph distinctly reveals that the ML decoder requires $2048^2$ operations compared to $160$ for the block-based FHT decoder. The number of operations needed by the ML decoder highly surpasses that of the block-FHT decoder even if the FHT-based decoders are executed sequentially. So the merits of decoding using FHT can be visibly perceived in terms of complexity. It is therefore advisable to employ this technique in scenarios where time sensitivity is crucial, particularly in {ultra-reliable low-latency communications} (URLLC) and industrial IoT applications.
%Time efficiency estimates depend on what we define to be a step.
%
% On peut nettement lire sur la graph, que le decodeur ML necessite $1024$ operations contre  $186$ pour le block-based FHT decoder. Le nombre d'operations requis par le decodeur ML est $5.5 \times$ superieur au block-FHT decoder. Meme si les  based FHT-based sont exexuté sequentiellment cette rapport ne se reduirait qu'à $3.5 \times$. Donc les mérite du decodage par FHT semble visible à la volet en terme de complexité.

\balance
\section{Conclusion}
This  paper presented low-complexity block-based encoding and decoding algorithms for short block length channels. In terms of the precise use-case, we were primarily concerned with the baseline 3GPP Short block transmissions in which payloads are encoded by Reed-Muller codes and  paired with orthogonal DMRS. In contemporary communication systems, the short block decoding often employs the utilization of DMRS-based least squares channel estimation, followed by maximum likelihood decoding. However, it is acknowledged that this approach can incur substantial computational complexity when processing long bit length codes.
We proposed an innovative approach to tackle this challenge by introducing the principle of block/segment encoding using First-Order RM Codes which  is amenable to low-cost decoding through block-based fast Hadamard transforms. The Block-based FHT has demonstrated to be cost-efficient with regards to decoding time, as it evolves from quadric to quasi-linear complexity with a manageable decline in performance. Additionally, by incorporating an adaptive DMRS/data power adjustment technique, we were able to bridge/reduce the performance gap and attained high sensitivity, leading to a good trade-off between performance and complexity to efficiently handle small payloads.
%\section*{Acknowledgment}

%The preferred spelling of the word ``acknowledgment'' in America is without
%an ``e'' after the ``g''. Avoid the stilted expression ``one of us (R. B.
%G.) thanks $\ldots$''. Instead, try ``R. B. G. thanks$\ldots$''. Put sponsor
%acknowledgments in the unnumbered footnote on the first page.

\vspace{12pt}

%The most popular algorithm for decoding first order RM codes is the algorithm based on the FHT [13, 14, 19].
%\color{red}
%IEEE conference templates contain guidance text for composing and formatting conference papers. Please ensure that all template text is removed from your conference paper prior to submission to the conference. Failure to remove the template text from your paper may result in your paper not being published.
%13. S.W. Golomb, ed., Digital Communications with Space Applications, Prentice-Hail, Englewood Cliffs, NJ(1964).
%14. R.R. Green, A serial orthogonal decoder, JPL Space ProgramsSummary, Vol. 37-39-IV (1966) pp. 247-253.
%19. E J. MacWilliams and N. J. A.Sloane, The Theory of Error-correcting Codes, North-Holland, Amsterdam, The Netherlands (1977).

\end{document}